# Single-Ended Fiber Latency Measurement with Picosecond-Accuracy Using Correlation OTDR

Michael H. Eiselt* and Annika Dochhan*
* ADVA Optical Networking SE, Maerzenquelle 1-3, 98617 Meiningen, Germany

**Abstract**—**Correlation OTDR with a bit rate of 10 Gbit/s is used to measure fiber latency with an accuracy of a few picoseconds. The concept is demonstrated measuring the chromatic dispersion of a 2.2-km fiber section.**

## I. Introduction

Synchronization in optical networks has recently become an important topic, as more and more time critical applications are transported in the network, and with the arrival of 5G networks the requirements in this area will become even more stringent. An application with a very tight requirement on synchronization is the transport of phase synchronous microwave signals over a multi-core fiber or a fiber bundle to feed a phase array antenna, as envisioned in the European BlueSpace project [1]. For a carrier frequency of 28 GHz, the differential latency is limited to approximately 3 ps. Other synchronization applications will require a knowledge of the transmission fiber latency with an accuracy of better than a nanosecond. A typical temperature dependence of the refractive index of optical fiber is on the order of $7 \cdot 10^{-6}$/degC [2]. Therefore, temperature changes of a 20-km fiber lead to a change of latency of approximately 700 ps/degC. Continuously monitoring the latency of the deployed fiber, of which usually only one end is accessible, will therefore become unavoidable in future latency sensitive networks.

A well-known method to measure an optical fiber length from one side is optical time-domain reflectometry (OTDR). However, as only a single pulse is used in every measurement run, a probe pulse width of 10 ns or more is typically required, limiting the accuracy to several nanoseconds. While coherent OTDR and other reflective methods can be used to achieve accuracies down to the picosecond range [3], these methods require narrow linewidth lasers, are quite complex and appear too costly to be deployed for network-wide fiber monitoring.

A variant of the pulse OTDR is the correlation OTDR, where an intensity modulated data sequence is sent into the fiber under test and a cross-correlation is performed between the back-reflected signal and the transmitted data sequence [4]. In this paper, we report on the use of correlation OTDR with a data rate of 10 Gbit/s to achieve a picosecond accuracy on the fiber latency, while requiring only a simple measurement setup.

## II. Experimental Evaluation

The setup to characterize a fiber of approximately 2.2 km length is shown in Figure 1. A 10-Gbit/s data stream was modulated onto an optical CW wave at 1550 nm, using a Mach-Zehnder modulator. The data stream had a period of 50 µs and consisted of a $2^7$-1 bit PRBS sequence, appended by a sequence of zeroes. With 50 µs the packet duration was more than twice the fiber round-trip time (RTT) of 22 µs, thus enabling the monitoring of signals having experienced multiple reflections.

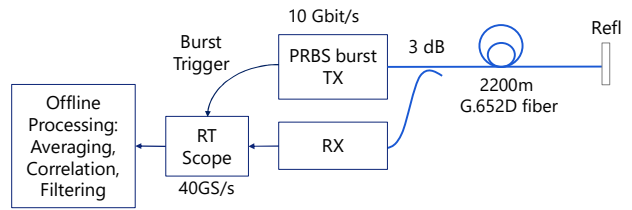

Fig. 1. Measurement setup: The 10 Gbit/s burst sequence was sent over 2200 m of G.652D fiber, which was terminated with a high reflectivity. The RX consisted of a 10G PIN/TIA, and the detected signal was captured by a 40-GS/s real-time oscilloscope.

The optical signal was sent into the fiber under test via a 3-dB coupler. To mark the beginning of the fiber, a small air gap was provided at the input connector between 3-dB coupler and transmission fiber. The end of the fiber was terminated with a high reflectivity to allow observing triple-reflection signals. Similar performance was achieved with an open (not angled) fiber end replacing the high reflectivity. The reflected signal was received after the 3-dB coupler and observed via a PIN/TIA receiver on a 40-GS/s real-time oscilloscope with a specified timing accuracy of 0.1 ppm. Data post-processing of 1000 captured traces was performed in Matlab. A typical received trace is shown in Figure 2 (left).

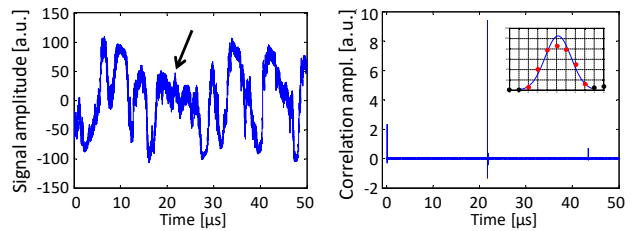

Fig. 2. Left: Typical received single reflection trace; Right: after averaging of 1000 traces and cross-correlation with transmitted PRBS sequence, inset shows fit of Gaussian pulse to one reflection peak.

The small peak around 22 µs (marked by the black arrow) corresponds to the reflected PRBS sequence from the end of the 2200-m fiber. The other signal variations, which a larger magnitude than the fiber-end reflection, can likely be attributed to coherent backscattering of the sequence of zeroes with limited extinction ratio. After averaging of the traces, a cross-correlation between the received signal and the transmitted PRBS sequence was performed. As only a single PRBS sequence was transmitted, surrounded by zeros, the cross-correlation contains pre- and post-cursors, which are mostly eliminated by filtering with a pre-calculated 255-tap filter. The result of the cross-correlation is shown in Figure 2 (right), where three peaks are distinguishable. In a final step, four parameters of a Gaussian shape (center, width, amplitude, offset) are fitted to each of the largest peaks in the cross-correlation to determine the center of the reflection points (see inset in Figure 2).

For the trace shown in Figure 2 (right), the (round-trip) latencies of the three main peaks are at 94.2372 ns, 21,733.1958 ns, and 43,372.1563 ns. Apparently, the third peak stems from triple-reflection of the optical burst at the fiber end, fiber input, and fiber end again and is therefore observed after the end of the fiber. This triple-reflection peak also points to the accuracy of the measurement. With the round-trip times of the fiber input and fiber end reflections, this peak should have occurred at 2 x 21,733.1958 ns – 94.2372 ns = 43,372.1544 ns, which differs by only 1.9 ps from the measured value.

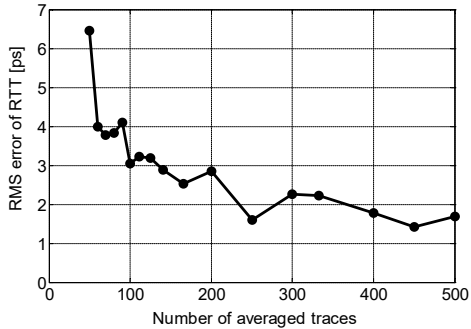

Fig. 3. RMS error of round-trip latency of the triple-reflection peak, as calculated from the fiber input and fiber end round-trip latencies.

While for these measurements 1000 traces were averaged, the measured traces were also re-evaluated with fewer averages. The recorded traces were divided into sub-sets, which were evaluated separately. The RMS error of the triple reflection latency as a function of the sub-set length, is shown in Figure 3. Already for 100 averages the round-trip measurement error is below 4 ps.

### III. APPLICATION FOR CHROMATIC DISPERSION MEASUREMENT

To further demonstrate the accuracy of the measurement method, the round-trip latency of the 2.2-km fiber with an open connector at the end was measured at seven different wavelengths. For each wavelength, 1000 traces were captured, and the time position of the input and output reflection peaks were determined by averaging and cross-correlation.

During the measurement, the temperature in the laboratory fluctuated slightly. With a temperature coefficient of $7 \cdot 10^{-6}$/degC, a temperature change of 1 degC would lead to a round-trip latency increase of 150 ps and therefore strongly impact the chromatic dispersion measurement. To evaluate the temperature dependent latency fluctuations, the measurement sets were divided into sub-sets of 250 traces, and the round-trip latency of the fiber-end peak was calculated for each sub-set of each wavelength measurement. Over the measurement time of 3.5 hours, a round-trip latency increase of 120 ps was observed, corresponding to a temperature increase of approximately 0.8 degC.

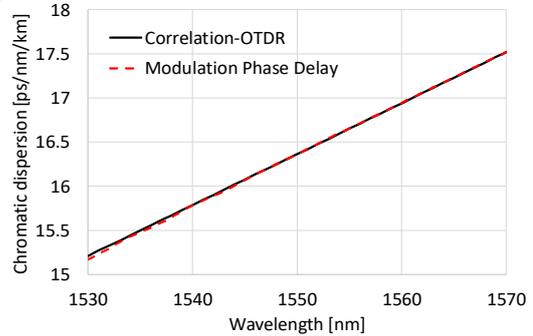

Fig. 4. Chromatic dispersion measured by new method (black solid line) and by modulation phase delay method (red dashed line).

For each wavelength, the average latency drift during the recording of 1000 traces was then compensated. The resulting latencies were fit to a 2nd-order polynomial of the wavelength with an RMS fitting error of 2.2 ps, and the chromatic dispersion in the C-band was calculated as shown in Figure 4. The results were compared to CD measurements using the modulation phase delay method, depicted as the dashed line in Figure 4. The maximum difference between these two measurements was 0.04 ps/nm/km over the C-band wavelength range.

### IV. SUMMARY

Correlation OTDR with a PRBS burst data rate of 10 Gbit/s was used to measure the latency of a 2.2-km fiber segment from one end. By observation of a triple reflection peak, the accuracy of the measurement was estimated to be better than a few picoseconds. Using the method with signals at different wavelengths yielded the chromatic dispersion of the fiber with a very good agreement with a modulation phase delay measurement. During all measurements, small drifts of the round-trip time were observed, which were attributed to temperature changes in the laboratory of less than one degC.

### ACKNOWLEDGMENT

This project has received funding from the European Union's Horizon 2020 research and innovation programme under grant agreement No 762055.